\documentclass{aa}  
\usepackage{graphicx}
\usepackage{txfonts}
\usepackage{lipsum}
\usepackage{subcaption}     
\usepackage{booktabs}
\usepackage{lscape}            
\usepackage{placeins}         
\usepackage{siunitx}
\usepackage[colorlinks=true,linkcolor=blue,citecolor=blue,urlcolor=blue]{hyperref}
\usepackage{array}
\usepackage{float}
\usepackage{amsmath}
\usepackage{multirow}

\begin{document}

\titlerunning{Binary-single scattering with ML}
\authorrunning{A. Farhani Asl et al.}

\title{Predicting the final states of binary-single scattering with machine learning}

\author{Ahmad Farhani Asl\inst{1}\corrauth{a.farhaniasl@gmail.com}
	\and David Fonseca Mota\inst{2}
	\and Dennis Fremstad\inst{2}
	\and Fatemeh Rahimi\inst{1} 
}

\institute{
	Department of Physics, Institute for Advanced Studies in Basic Sciences (IASBS), Zanjan, Iran
	\and Institute of Theoretical Astrophysics, University of Oslo, Oslo, Norway
}

\date{Received August 12, 2026}

\abstract
{Binary-single encounters are particularly frequent in dense stellar environments, where they play a central role in shaping the dynamical evolution of their host systems. However, predicting their final outcomes remains an open question due to the intrinsic chaotic nature of the three-body problem. This challenge motivates the adoption of data-driven machine learning (ML) methods.}
{We investigate whether ML can predict the final outcomes of binary-single encounters from initial conditions alone.}
{We generated $5.8\times10^6$ binary-single scattering simulations using the REBOUND $N$-body package with the IAS15 integrator. A cascaded binary classification strategy, comprising four sequential XGBoost classifiers, and a single multi-class model were trained on the synthetic dataset and compared.}
{The cascaded strategy outperforms the single multi-class model across all metrics. F1-scores for the cascaded models exceed $0.92$, with precision-recall area under the curve (PR-AUC) values reaching $0.99$, compared to $0.95$ for the multi-class model. Feature importance analysis identifies encounter timescale, binary hardness, and mass ratio as key predictors. Misclassification analysis shows that prediction failures concentrate near chaotic boundaries where the outcome is sensitive to small perturbations. Speed benchmarks demonstrate that the cascaded models are up to $300$ times faster than direct $N$-body integrations. However, all models fail to generalize to new datasets, highlighting a key limitation.}
{This study demonstrates that, for any specific environment and data distribution, the proposed cascaded ML strategy provides a robust and rapid framework for predicting binary-single scattering outcomes.}


\maketitle

\nolinenumbers

\section{Introduction}

With only two components, binaries are the smallest stellar systems in the Universe, yet they constitute a significant portion of larger structures. Approximately one-third of stars in the Galactic field are estimated to reside in binary systems \citep{Raghavan2010}. Open clusters, however, exhibit significantly higher binary fractions, reaching up to $70\%$ \citep{Sollima2010}. In more compact stellar systems, such as globular clusters, where central densities can exceed $10^5 M_\odot \, \mathrm{pc}^{-3}$ \citep{Calura2019}, the binary fraction varies from a few percent up to $10\%$ \citep{Ji2015}. Despite this relatively lower abundance of binaries, the sufficiently high density ensures frequent significant gravitational interactions between binaries and the field single stars \citep{Henon1975, Sigurdsson1993}. The consequences of such binary-single encounters are not limited to the interacting bodies themselves, but cumulatively drive the thermodynamic evolution of the host stellar systems \citep{Hut1983a}. 

According to \citet{Heggie1975}, a binary is classified as either hard or soft depending on whether its binding energy exceeds or falls below the mean kinetic energy of the surrounding field stars. Hard binaries tend to become harder, i.e. tighter, with each gravitational interaction, transferring energy to the stellar background and heating the ambient stars. Soft binaries, in contrast, absorb energy and gradually soften until eventual complete disruption. This continuous energy flow, as observed in globular clusters, regulates mass segregation \citep{Spitzer1975}, core oscillations \citep{Bettwieser1984, Hurley2012}, and the formation of exotic objects such as blue stragglers and compact X-ray binaries \citep{Zwart2004, Ferraro2009}. In addition, within the concentrated black hole subsystems in globular clusters, binary-single interactions can become extreme, progressively hardening binaries while ejecting single black holes and thereby triggering a gradual self-depletion process \citep{Shirazi2024}. Beyond merely dynamical effects, recent studies show that resonant binary-single interactions can produce gravitational wave (GW) transient signatures \citep{Codazzo2024, Rando2025}. Given the considerable impact of these encounters, the dynamical evolution of individual binary-single scatterings and the variety of their final states have been recently the subject of extensive theoretical and numerical investigation \citep[e.g.,][and references therein]{Ginat2021a, Hellstrom2022, Wang2025, Heinze2025}.

An individual binary-single encounter represents a realization of the classical chaotic three-body problem. The relative likelihood of the outcomes, for instance whether the binary survives, traps the intruder star, or dissolves, depends sensitively on the initial conditions. Earlier studies suggested that binary hardness, mass ratio, and encounter velocity play key roles in determining the encounter outcomes \citep{McMillan1996, Heggie1993}. Despite decades of analytical and numerical studies, predicting the final state of an individual encounter remains unanswered because the scattering process is fundamentally non-integrable \citep{Valtonen2005, Stone2019}. Minute perturbations in the initial configuration can evolve into qualitatively different dynamical outcomes, particularly during resonant phases, in which the three bodies undergo multiple temporary captures before the system decays into a permanent final state \citep{Zwart2023}. This intrinsic sensitivity limits the practical predictability of deterministic approaches and motivates the use of data-driven statistical methods capable of learning probabilistic mappings between initial conditions and final state classes.

Machine learning (ML) methods are increasingly utilized in dynamical systems where large numbers of numerical integrations are computationally expensive. In the context of the three-body problem, supervised learning algorithms can, in principle, identify patterns from simulation data and use them to predict final states without performing N-body integrations. Recent studies have demonstrated the effectiveness of tree-based ensemble models in gravitational dynamics. For example, \citet{Neto2024} trained an XGBoost classifier on $10^5$ IAS15 simulations of the restricted three-body problem, using the initial conditions as input features to predict orbital stability, achieving an accuracy exceeding $98\%$. Similarly, \citet{Pinheiro2024} applied XGBoost to predict stability in planetary systems using orbital parameters such as mass ratios, eccentricities, semi-major axes, and inclinations as features, also achieving high predictive accuracy. These studies indicate that tree-based ensemble methods, including Random Forest, Gradient Boosting, and XGBoost, are well suited for problems with strongly nonlinear and heterogeneous parameter spaces.

In this work, we trained a series of ML models to predict the final state of a binary-single encounter from its initial conditions, comparing a cascaded binary classification strategy against a single multi-class model. Trained on $5.8\times10^6$ simulations performed via the REBOUND $N$-body package, the cascaded models demonstrate high accuracy (with precision-recall area under the curve, PR-AUC, and receiver operating characteristic area under the curve, ROC-AUC, reaching $0.99$) and outperform the multi-class model. We find that the encounter timescale, binary hardness, and mass ratio are the top important features across models. Misclassification analyses reveal the chaotic boundaries where models struggle the most, providing additional insight beyond feature importance. Speed benchmarks show that our trained models are up to $300$ times faster than direct $N$-body integrations, making them promising for population synthesis and cluster studies. However, our models fail to generalize to new datasets, highlighting a key limitation.

The remainder of this paper is organized as follows. Section~\ref{sec:methods} describes the numerical setup, data generation, classification criteria, and ML framework. Section~\ref{sec:results} presents the model performance, feature importance, misclassification analysis, and speed benchmark. Section~\ref{sec:discussion} interprets the results, while Sect.~\ref{sec:future_works} addresses the limitations and outlines future directions. Section~\ref{sec:conclusions} summarizes the main findings and concludes the paper.

\section{Method}
\label{sec:methods}

\subsection{Initial relative distance and velocity}
\label{sec:initial_seperation}

\begin{figure}[h!]
	\centering 
	\includegraphics[width=0.8\hsize]{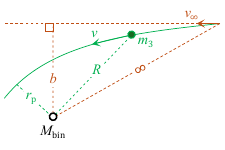}	
	\caption{Initial geometry of a binary-single encounter.}
	
	\label{fig:hyperbola}
\end{figure}

We began by specifying the initial distance, $\vec{R}$ (Fig.~\ref{fig:hyperbola}), between the binary system (masses $m_1$ and $m_2$) and the intruder single body ($m_3$). As noted by \citet{Heggie2003}, choosing this distance involves a trade-off. Overly large values of $\vec{R}$ incur prohibitive computational costs, whereas values that are too small risk introducing unphysical perturbations that compromise the reliability of the simulations. To address these competing constraints, we adopted a criterion limiting the intruder's tidal perturbation to $10^{-3}$ of the binary's internal gravitational acceleration. Based on earlier test runs, this threshold approximately ensures that the intruder is initialized outside the binary cross section, while balancing computational cost against energy conservation accuracy. The tidal acceleration scales as
\[
a_{\rm tidal} \sim \frac{2\, G\, m_3}{{R}^3}\, a,
\]
where $G$ is the gravitational constant and $a$ is the semimajor axis of the binary. This expression is derived by applying a Taylor expansion to the intruder's gravitational acceleration exerted on each binary component, $G m_3 (\vec{R} + \vec{a}) / |\vec{R} + \vec{a}|^3$, to first order in $a/R \ll 1$, and subtracting the corresponding acceleration on the binary center of mass (COM), i.e., $G m_3 \vec{R} / R^3$. The binary's internal acceleration is
\[
a_{\rm bin} \sim \frac{G M_{\rm bin}}{a^2},
\]
where $M_{\rm bin} = m_1 + m_2$. Requiring $a_{\rm tidal} / a_{\rm bin} \lesssim 10^{-3}$ yields
\begin{equation}
\label{eq:initial_R}
R \gtrsim 10\, \left(\frac{2\,m_3}{M_{\rm bin}}\right)^{1/3} a.
\end{equation}
We therefore initialized the intruder single body position at the minimum value of Eq.~\ref{eq:initial_R} relative to the binary COM. As shown in Fig.~\ref{fig:hyperbola}, we computed the initial relative velocity, $\vec{v}$, corresponding to the asymptotic velocity, $\vec{v}_\infty$, and impact parameter, $b$, using the hyperbolic trajectory of the two-body system (consisting of the binary COM and the intruder body) under conservation of energy and angular momentum.

\subsection{Binary initial configuration}
\label{sec:binary_setup}

\begin{figure}[h!]
	\centering 
	\includegraphics[width=0.6\hsize]{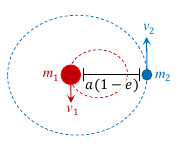}	
	\caption{Initial configuration of a binary system.}
	
	\label{fig:binary}
\end{figure}

Figure~\ref{fig:binary} shows the initial configuration of the binary system, when the intruder body is at $\vec{R}$ relative to the binary COM. The binary components were placed on the $x$-axis at their pericenter phase. Their velocity vectors, $\vec{v}_1$ and $\vec{v}_2$, were set along the $y$-axis, such that the binary components orbit counterclockwise in the $x$-$y$ plane, with the angular momentum vector pointing toward the $+z$-axis. The semimajor axis, $a$, and eccentricity, $e$, determine the magnitudes of the velocities. 

\subsection{Initial parameters}

Each simulation was initialized with nine principal parameters that fully specify the configuration of a binary-single encounter. These parameters, listed in Table~\ref{tab:initial_params} along with their sampling ranges and distributions, include the three body masses $m_i$ ($i=1,2,3$), the binary semimajor axis $a$, the binary eccentricity $e$, the impact parameter $b$, and the asymptotic relative velocity $v_\infty$ with a direction defined by polar and azimuthal angles, $\theta$ and $\phi$.  

Stellar masses are distributed between $0.08$ and $150 M_{\odot}$ with $m_2 \le m_1$ to avoid duplicate configurations, covering the typical mass range found in globular and open clusters, from low-mass M dwarfs to massive O-type stars and black holes \citep{Kroupa2001, Chabrier2003, Morscher2015}. The binary semi-major axis, $a$, was sampled between $0.01$ and $10^5$ au, spanning extremely wide binaries \citep{Makarov2025}, while remaining well within the nonrelativistic regime. Even for the most extreme configuration in our parameter space (minimum $a$, maximum $e$, $m_1$, and $m_2$), the pericenter distance is $10^{-5}$ au, which exceeds the Schwarzschild radius of approximately $6\times10^{-6}$ au. Eccentricities vary from 0 to 0.999, producing circular to highly eccentric orbits, the latter being of particular interest for GW studies \citep{Samsing2018, Rodriguez2018}. The squared impact parameter $b^2$, corresponding to the encounter cross section, is sampled from 0 to $25r_{\rm crit}$, where $r_{\rm crit} = R/2$ (see Eq.~\ref{eq:initial_R}), covering the full transition from head-on to distant encounters. The single body asymptotic velocity $v_{\infty}$ is sampled between $10^{-5}$ and $10^{3}$ au\,yr$^{-1}$ with respect to the binary COM, spanning across the velocity dispersions of various environments, from the cores of globular clusters \citep{Harris1996, Leanza2024} to the nuclear star clusters of galaxies \citep{Schodel2009}. To ensure isotropic coverage of approach directions, $\cos\theta$ and $\phi$ are sampled uniformly in $[-1,1]$ and $[0,2\pi]$, respectively. Here $\theta$ is the angle between the intruder's approach direction and the binary angular momentum vector ($+z$-axis), and $\phi$ is the azimuthal angle measured from the binary pericenter direction ($+x$-axis). This is important because the final state of binary-single encounters depends sensitively on the mutual inclination between the binary orbital plane and the intruder trajectory \citep{Antognini2016}.

\begin{table}[h!]
	\caption{Initial parameter ranges and sampling distributions.}
	\label{tab:initial_params}
	\centering
	\begin{tabular}{lp{3cm}l}
		\toprule\toprule
		Parameter & Range & Distribution \\
		\midrule
		$m_i$ & $[0.08,\;150]$ $M_{\odot}$ & Log-uniform \\
		$a$ & $[10^{-2},\;10^{5}]$ au & Log-uniform \\
		$e$ & $[0.0,\;0.999]$ & Uniform \\
		$b^2$ & $[0,\;25 r_\mathrm{crit}]$ & Uniform \\
		$v_{\infty}$ & $[10^{-5},\;10^{3}]$ au\,yr$^{-1}$ & Log-uniform \\
		$\cos\theta$ & $[-1,\;1]$ & Uniform \\
		$\phi$ & $[0,\;2\pi]$ & Uniform \\
		\bottomrule
	\end{tabular}	
\end{table}

\subsection{Data generation}
\label{sec:code}

We used the REBOUND $N$-body package \citep{Rein2012} to generate a synthetic dataset of binary-single encounter simulations. Among the integrators available in the REBOUND suite, we selected IAS15 \citep{Rein2015}, a 15th-order adaptive integrator that conserves energy to machine precision and reliably handles close encounters. IAS15 is widely used in three-body simulations across a range of applications, including planetary stability, asteroid dynamics, and exoplanetary systems \citep{Jaworska2026, Smirnov2025, Kenger2025}. Alternative integrators available in REBOUND, though faster, are either less accurate or demand case-specific parameter tuning, which introduces unnecessary risks and complexities \citep{Wisdom1991, Rein2015WHFAST, Rein2019}. All simulations were performed using $G=4\pi^2$, expressed in astronomical units (au), solar masses ($M_\sun$), and years (yr). The simulations were executed in parallel across multiple CPU cores.

All simulations were terminated at $t_\mathrm{max} = 3\,t_{\rm enc}$, where $t_{\rm enc} = R / v$, regardless of whether the system had reached a permanent final state. This fixed cutoff was chosen to eliminate time dependence from the classification while ensuring a uniform prediction horizon across all encounters. Our previous study \citep{Farhani2026}, using the same termination criterion, found that only a few per cent of the final states remained transient, confirming that this timescale is sufficient for the vast majority of encounters to settle into long-lived configurations. This choice is further supported by the expectation that resonant three-body interactions typically complete within a few crossing times before the system reaches a well-defined final state \citep{Heggie1993, Samsing2014}. This is consistent with the chaotic three-body dynamics, where the system alternates between long resonant episodes and rapid decay \citep{Shevchenko2010}. In total, we generated $5.8\times10^6$ independent binary-single scattering simulations, covering the full parameter space described in Table~\ref{tab:initial_params}. 

\subsection{Classification of final states}
\label{sec:outcome_classification}

Upon completion of each simulation, the system was classified into one of four final-state classes, namely ionization, flyby, exchange, or hierarchical triple (hereafter hierarchy for brevity). We classified the dynamical outcomes using three complementary diagnostics: (i) the pairwise specific energies of the three bodies, (ii) the velocity of each body relative to the escape velocity from the COM of the other two bodies, and (iii) the direction of the relative velocity between a potential escapee and the remaining bound pair. The pairwise specific energy of stars $i$ and $j$ is defined as
\begin{equation}
	\label{eq:specific_energy}
	E_{ij} = \frac{1}{2} v_{ij}^2 - \frac{G (m_i + m_j)}{r_{ij}},
\end{equation}
where $v_{ij}$ and $r_{ij}$ are the relative velocity and distance, respectively. We adopted pairwise specific energies rather than binding energies to avoid mass-induced biases, as binding energy can overlook tight binaries with low-mass components, whereas specific energy provides a purely geometric, mass-invariant criterion. 

To determine whether a body has escaped the system, we computed its velocity relative to the COM of the other two bodies. The escape velocity of body $k$ relative to the pair $(i,j)$ is
\begin{equation}
	\label{eq:escape_velocity}
	v_{\mathrm{esc},k} = \sqrt{\frac{2G (m_i+m_j)}{|\vec{r}_k - \vec{r}_{\mathrm{com},ij}|}}.
\end{equation}
A body was considered unbound if its relative speed $|\vec{v}_k - \vec{v}_{\mathrm{com},ij}|$ exceeds $v_{\mathrm{esc},k}$ (Eq.~\ref{eq:escape_velocity}). We additionally verified that the single body is moving away from the COM of the pair by checking the sign of the relative radial velocity
\begin{equation}
	\label{eq:velocity_direction}
	(\vec{r}_i - \vec{r}_{{\rm com},jk}) \cdot
	(\vec{v}_i - \vec{v}_{{\rm com},jk}) > 0 ,
\end{equation}
where $\vec{r}_i$ and $\vec{v}_i$ are the position and velocity of the body under consideration, and $\vec{r}_{\mathrm{com},jk}$ and $\vec{v}_{\mathrm{com},jk}$ are the position and velocity of the COM of the remaining pair, respectively.

\begin{figure}[h!]
	\centering
	\includegraphics[width=\linewidth]{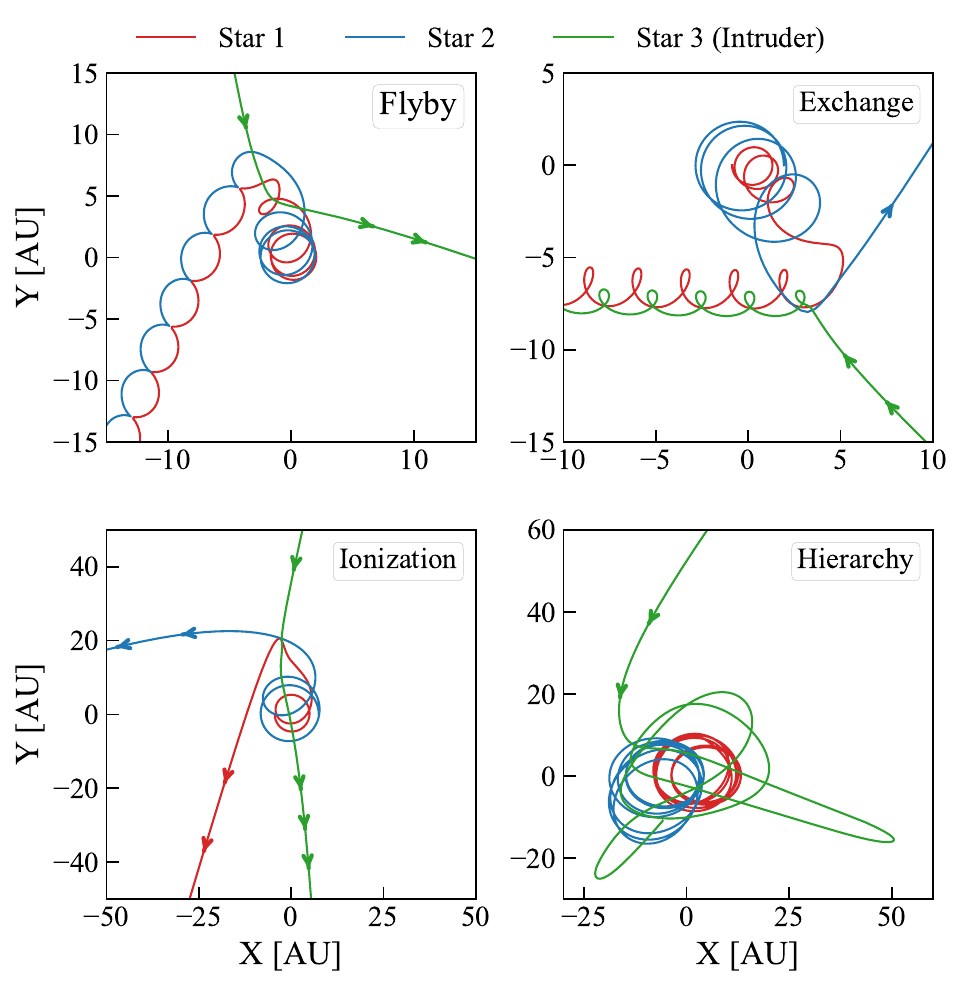}
	\caption{Example trajectories of the four possible outcomes of binary-single encounters in the binary COM frame. Each panel shows a representative simulation from our dataset. Colors distinguish the three bodies: original binary components (Star 1 in red, Star 2 in blue) and the intruder (Star 3 in green). Arrows indicate the direction of motion, highlighting the incoming and escaping stars.}
	\label{fig:outcome_examples}
\end{figure}

The final state was classified as follows. If no pair satisfied $E_{ij}<0$ (Eq.~\ref{eq:specific_energy}), the final state was labeled as ionization, in which all three bodies are unbound relative to each other. If exactly one pair was bound ($E_{ij}<0$ for a specific combination $ij$) and the remaining body satisfied the escape conditions of Eqs.~\ref{eq:escape_velocity} and \ref{eq:velocity_direction}, the case was considered either a flyby (if the bound pair was composed of $m_1$ and $m_2$) or an exchange (if the bound pair included the intruder body $m_3$). Depending on which component of the original binary was replaced by the intruder body, we further classified exchanges into two subtypes, exchange1 and exchange2, corresponding to the ejection of $m_1$ and $m_2$, respectively. Finally, the system was classified as hierarchy when the outer body was captured by the most bound pair's COM (i.e., Eq.~\ref{eq:escape_velocity} was not satisfied) or when more than one pair was found bound. The stability of the formed hierarchies was assessed using the criteria of \citet{Mardling2001} and \citet{Vynatheya2022}, where only one case in the entire dataset satisfied the former, consistent with the expected rarity of long-lived stable hierarchies. All remaining unclassified cases, accounting for less than $0.01\%$ of the dataset, were excluded to ensure clean class definitions. The classification procedure yielded the following class distribution: flyby $46.8\%$ ($2\,708\,809$), exchange $27.5\%$ ($1\,591\,772$), hierarchy $24.8\%$ ($1\,437\,242$), and ionization $0.85\%$ ($47\,630$). Representative cases of the four final states from our dataset are shown in Fig.~\ref{fig:outcome_examples}. 

We emphasize that the final-state classification is based solely on the conditions at the last snapshot of each simulation. Predicting the instantaneous final state is more physically relevant in dense environments, where the system is unlikely to remain isolated shortly after the encounter, and neighboring bodies may intervene before the configuration can relax \citep{Fregeau2006, Geller2015}. Therefore, we do not perform extended integrations to assess long-term stability, as this is of secondary importance for the present study.

\subsection{Data preprocessing}

Prior to training, we removed inaccurate cases, excluding simulations with relative energy errors exceeding $10^{-8}$, final eccentricities above $0.9999$ \citep{Mortari2007}, and final semimajor axes outside the initial sampling range ($a \in [0.01, 10^5]$ au). Although this relative energy error threshold is more stringent than the $10^{-1}$ tolerance suggested by \citet{Zwart2014}, we chose this conservative criterion to ensure compatibility with high-precision codes such as NBODY6++GPU and PeTar \citep{Wang2015, Wang2020}.  

\subsection{Feature vector construction}
\label{sec:features}

The choice of input features for the ML model was guided by physical principles established in the literature on binary-single scattering (Table~\ref{tab:features}). The hardness parameter $\mathcal{H}$ is the primary statistical determinant of binary evolution, encoding the direction of energy flow in dense environments \citep{Heggie1975, Ginat2021a}. Energetic parameters, including the Sofronov number $\Theta$, the energy penetration parameter $\eta$, and the focusing variables $\beta$ and $\mathcal{F}$, collectively characterize the encounter's ability to disrupt the binary and the degree to which gravitational focusing enhances the interaction cross section \citep{Samsing2014, Samsing2018}. The mass ratios $q_2$ and $q_3$, mass asymmetry $\mu_a$, mass hierarchy $\mu_h$, and mass entropy $\delta_m$ encode the level of mass variance among the three bodies, which influences the branching likelihood between exchange and hierarchy classes \citep{Heggie1996, Antognini2014}. Geometric parameters define the encounter configuration and timescale. The cross section ratio $(b/a)^2$, scaled velocity $\upsilon$, encounter timescale $\tau$, and angular momentum ratio $\mathcal{L}$ determine the proximity, duration, and orbital geometry of the interaction, all of which strongly influence the final state classification \citep{Heggie1993, Antognini2016}.

These physically motivated features provide a comprehensive representation of the initial state. Prior to training, all features were standardized to zero mean and unit variance using scikit-learn's built-in tools, which improves model accuracy and generalization \citep{Pedregosa2011, Gelman2008}.

\begin{table}[h!]
	\centering
	\caption{Features used for training the ML models.}
	\label{tab:features}
	\begin{tabular}{ll}
		\toprule\toprule
		Feature & Definition \\
		\midrule
		Hardness ($\mathcal{H}$) & $G m_1 m_2 / (a \, \mu \, v_\infty^2)$ \\
		Binary type ($\mathcal{B}$) & hard or soft \\
		Binary mass ratio ($q_2$) & $m_2 / m_1$\\
		System mass ratio ($q_{3}$) & $m_3 / M_{\rm bin}$ \\
		Sofronov number ($\Theta$) & $2 G M_{\rm bin} / (a v_\infty^2)$ \\
		Gravitational focusing ($\mathcal{F}$) & $G M_{\rm bin} / (b v_\infty^2)$ \\
		Impact focusing ($\beta$) & $b v_\infty^2 / (G M_{\rm bin})$ \\
		Energy penetration ($\eta$) & $m_3 v_\infty^2 / |E_{\rm bin}|$ \\
		Binary energy fraction ($\epsilon$) & $U_{\rm bin} / U_{\rm tot}$ \\
		Mass asymmetry ($\mu_{\rm a}$) & $(m_{+} - m_{-}) / (m_{+} + m_{-})$ \\
		Mass hierarchy ($\mu_{\pm}$) & $m_{+} / m_{-}$ \\
		Mass entropy ($\delta_{\rm m}$) & $|\sum_i (m_i / M_{\rm tot}) \ln(m_i / M_{\rm tot})|$ \\
		Scaled cross section ($\mathcal{A}$) & $(b / a)^2$ \\
		Scaled initial velocity ($\upsilon$) & $v / \sqrt{G(m_1+m_2)/a}$ \\
		Scaled encounter time ($\tau$) & $t_{\rm enc} / P$ \\
		Lensing parameter ($\mathcal{R}$) & $r_{\rm p} / b$ \\
		Scaled angular momentum ($\mathcal{L}$) & $(m_3 / M_{\rm bin}) \, b v_{\infty} / (a v_{\rm orb})$ \\
		\bottomrule
	\end{tabular}
	\tablefoot{For parameters not explicitly defined in the text, $\mu = m_3 M_{\rm bin} / M_{\rm tot}$ denotes the reduced mass of the intruder-binary system, with $M_{\rm tot} = m_1 + m_2 + m_3$. The binary binding energy is $E_{\rm bin} = (m_1 v_1^2 + m_2 v_2^2)/2 - G m_1 m_2 / a$, where $v_1$ and $v_2$ are the velocities of the binary components relative to the system center of mass. $U_{\rm bin}$ and $U_{\rm tot}$ represent the potential energy of the binary and the entire system, respectively. $m_+$ and $m_-$ are the most and least massive bodies. $P = 2\pi \sqrt{a^3 / (G M_{\rm bin})}$ is the Keplerian orbital period of the binary, $r_{\rm p}$ is the encounter pericenter distance (see Fig.~\ref{fig:hyperbola}), and $v_{\rm orb} = \sqrt{G M_{\rm bin} / a}$ is the binary orbital velocity.}
\end{table}

\subsection{Model training}
\label{sec:model}

\begin{figure}[h!]
	\centering 
	\includegraphics[width=0.85\hsize]{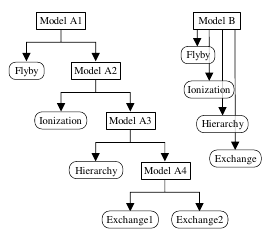}	
	\caption{Illustration of the two classification strategies employed in this study. The cascaded sequence of four binary classifiers (A1–A4, left) and the single general multi-class model (B, right).}	
	\label{fig:model_diagram}
\end{figure}

We employed two complementary classification strategies to systematically address the multi-class consequential nature of binary-single encounters. First, as illustrated in Fig.~\ref{fig:model_diagram}, we adopted a tree-based (cascaded) approach organized into four sequential binary classifiers. Model A1 distinguishes flybys, where binary is perturbed but survives, from all disruptive outcomes (ionization, exchange, and hierarchy). Model A2 identifies ionization events among the remaining non-flyby cases. Model A3 separates hierarchies from exchanges. Finally, Model A4 predicts the exchange subtypes (exchange1 and exchange2). Second, for completeness and comparison, we also trained a single general multi-class Model B to simultaneously classify all four outcomes in a single attempt.

We trained all models using XGBoost \citep{Chen2016}, a gradient-boosted ensemble of decision trees, chosen for its ability to capture highly nonlinear decision boundaries and its strong performance on structured tabular data \citep{Albertyn2025}. Following \citet{Chen2004, Chen2025}, we randomly undersampled the majority class to create a balanced dataset with a 1:1 class ratio for each model individually. Weighted sampling strategies were also tested but yielded poor predictive performance for minority classes. This is likely due to the chaotic nature of binary-single interactions, where assigning higher weights to rare outcomes does not compensate for the lack of representative examples in the training data. However, undersampling forces the model to learn from equal numbers of each class, thereby improving generalization. The data were then split into training and test sets using an 80/20 ratio, following the recommended practice of \citet{Gholamy2018}. 

\begin{table}[h!]
	\caption{Optimized hyperparameters for each model.}
	\label{tab:hyperparams}
	\centering
	\begin{tabular}{l l l l l l}
		\toprule\toprule
		\multirow{3}{*}{Parameter} & \multicolumn{5}{c}{Model} \\
		\cmidrule(lr){2-6}
		& A1 & A2 & A3 & A4 & B \\
		\midrule
		n\_estimators & $2000$ & $1951$ & $2000$ & $1580$ & $2000$ \\
		max\_depth & $24$ & $10$ & $22$ & $20$ & $20$ \\
		learning\_rate & $0.01$ & $0.1$ & $0.01$ & $0.03$ & $0.008$ \\
		subsample & $1.0$ & $0.8$ & $1.0$ & $1.0$ & $1.0$ \\
		colsample\_bytree & $0.92$ & $0.8$ & $0.84$ & $1.0$ & $1.0$ \\
		min\_child\_weight & $5$ & $1$ & $1$ & $1$ & $1$ \\
		gamma & $0.044$ & $0.1$ & $0.0$ & $0.0$ & $0.0$ \\
		reg\_alpha & $0.26$ & $0.35$ & $1.0$ & $1.0$ & $0.0$ \\
		reg\_lambda & $1.5$ & $1.48$ & $0.5$ & $1.5$ & $0.84$ \\
		\bottomrule
	\end{tabular}
\end{table}

For each model, hyperparameters were optimized via Bayesian search on a subset ($25\%$) of the entire dataset \citep{Kenny2024, Brochu2010}. As shown in Table~\ref{tab:hyperparams}, the optimized hyperparameters vary across models, reflecting the differing complexities of the decision boundaries at each classification level. Notably, Model A2 required the shallowest trees (max\_depth $= 10$) and the highest learning rate ($0.1$), suggesting a relatively simpler separation of the parameter space, whereas Model B adopted the lowest learning rate ($0.008$) and a moderate tree depth, consistent with the increased difficulty of the multi-class problem. Model performance was assessed using eight standard metrics: accuracy, precision, recall, F1-score, ROC-AUC, PR-AUC, Brier score, and expected calibration error (ECE). To ensure a robust evaluation, we additionally performed ten-fold cross-validation \citep{Hastie2009, Liu2023}. Training was carried out on an NVIDIA RTX $2070$ Super GPU using the XGBoost Python library.

\section{Results}
\label{sec:results}

\subsection{Classification performance}

The performance metrics for all five models are summarized in Table~\ref{tab:metrics}. Models A1, A2, and A4 attain near-perfect performance, all with ROC-AUC and PR-AUC above $0.99$. Model A2 delivers the best overall classification, with almost all metrics exceeding $0.99$. Model A3 exhibits marginally lower performance for the hierarchy class (F1-score $= 0.92$), with ROC-AUC and PR-AUC of around $0.98$. For comparison, the cascaded binary classifiers (Models A1 to A4) yield higher scores compared to Model B across all classes. The hierarchy class proves the most challenging for Model B (F1-score $= 0.83$, ROC-AUC $= 0.980$, PR-AUC $= 0.946$), while ionization remains relatively well predicted (F1-score $= 0.95$), following the same trend observed in the cascaded models. The top-2 accuracy of Model B is $0.986$, indicating that the true class is almost invariably ranked among the two highest probability predictions. The saved model sizes (in JSON format) are $860$ MB (A1), $6$ MB (A2), $4.6$ GB (A3), $770$ MB (A4), and $1.5$ GB (B).

\begin{table}[h!]
	\caption{Performance metrics of the trained models.}
	\label{tab:metrics}
	\centering
	\begin{tabular}{l c c c c c c}
		\toprule\toprule
		Class & Acc. & Pre. & Rec. & F1 & ROC & PR \\
		\midrule
		\multicolumn{7}{c}{Model A1} \\
		\midrule
		Flyby & 0.96 & 0.97 & 0.96 & 0.97 & \multirow{2}{*}{$0.995$} & \multirow{2}{*}{$0.996$} \\
		Disruption & 0.97 & 0.96 & 0.97 & 0.97 & & \\
		\midrule
		\multicolumn{7}{c}{Model A2} \\
		\midrule
		Ionization & 0.99 & 0.98 & 0.99 & 0.99 & \multirow{2}{*}{$0.996$} & \multirow{2}{*}{$0.997$} \\
		Other & 0.98 & 0.99 & 0.98 & 0.98 & & \\
		\midrule
		\multicolumn{7}{c}{Model A3} \\
		\midrule
		Hierarchy & 0.90 & 0.95 & 0.90 & 0.92 & \multirow{2}{*}{$0.978$} & \multirow{2}{*}{$0.981$} \\
		Exchange & 0.95 & 0.90 & 0.95 & 0.93 & & \\
		\midrule
		\multicolumn{7}{c}{Model A4} \\
		\midrule
		Exchange1 & 0.95 & 0.96 & 0.95 & 0.95 & \multirow{2}{*}{$0.992$} & \multirow{2}{*}{$0.992$} \\
		Exchange2 & 0.96 & 0.95 & 0.96 & 0.95 & & \\
		\midrule
		\multicolumn{7}{c}{Model B} \\
		\midrule
		Flyby & 0.89 & 0.96 & 0.89 & 0.92 & \multirow{4}{*}{$0.980$} & \multirow{4}{*}{$0.946$} \\
		Ionization & 0.97 & 0.92 & 0.97 & 0.95 & & \\
		Hierarchy & 0.80 & 0.85 & 0.80 & 0.83 & & \\		
		Exchange & 0.88 & 0.83 & 0.88 & 0.85  & & \\		
		\bottomrule
	\end{tabular}
	\tablefoot{From left to right, columns indicate class labels, accuracy, precision, recall, F1-score, ROC-AUC, and PR-AUC.}
\end{table}

The Brier scores are consistently below $0.06$ for all cascaded models, indicating well calibrated probability estimates, while Model B shows slightly higher Brier scores for identifying exchange ($0.057$) and hierarchy ($0.064$). For all cascaded models, ECE remains below $0.02$, while Model B exhibits larger values, particularly for hierarchy ($0.037$) and exchange ($0.035$). The ten-fold cross-validation standard deviations of the accuracies are below $0.002$ for all models.

\subsection{Feature importance}

\begin{figure}[h!]
	\centering
	\includegraphics[width=0.9\linewidth]{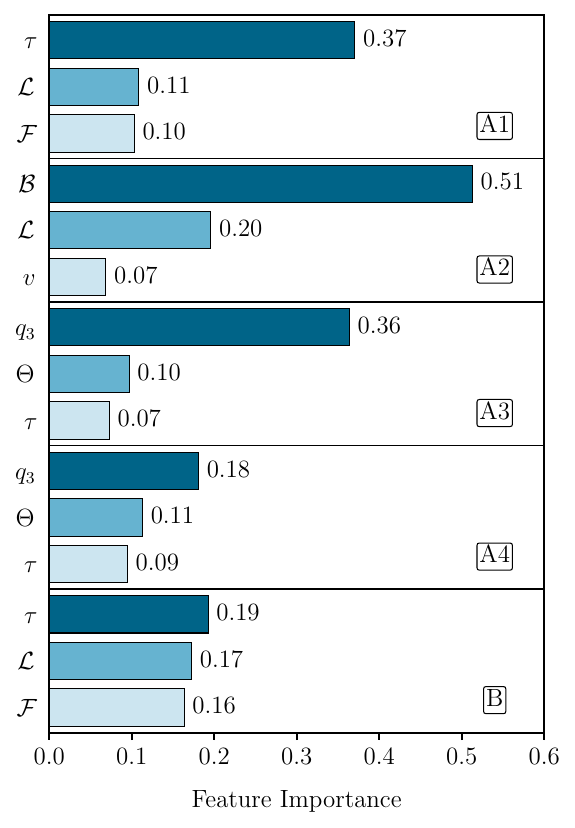}
	\caption{Top three important features for each of the five trained models (A1 to A4 and B). The bars represent the relative contribution of each feature to the model's predictions, with importance values normalized to sum to unity per model.}
	\label{fig:feature_importances}
\end{figure}

The distinct physical parameters governing each classification decision are revealed by the top three important features for each model, as shown in Fig.~\ref{fig:feature_importances}. For Model A1, the encounter timescale ($\tau = t_{\rm enc}/P$) is the dominant feature, followed by the scaled angular momentum ($\mathcal{L}$) and gravitational focusing ($\mathcal{F}$). Model A2 relies predominantly on the binary type ($\mathcal{B}$), contributing over half of the predictive power, with ($\mathcal{L}$) comes next of $20\%$ importance. In Model A3, the system mass ratio ($q_3$) and Sofronov number $\Theta$ are the most influential features with $36\%$ and $10\%$ importances, respectively. Model A4 similarly depends on $q_3$ ($18\%$), followed by $\Theta$ ($11\%$) and $\tau$ ($9\%$). For Model B, the top features are $\tau$ ($19\%$), $\mathcal{L}$ ($17\%$), and $\mathcal{F}$ ($16\%$).

\subsection{Misclassifications and chaotic boundaries}

\begin{figure}[h!]
	\centering
	\includegraphics[width=\linewidth]{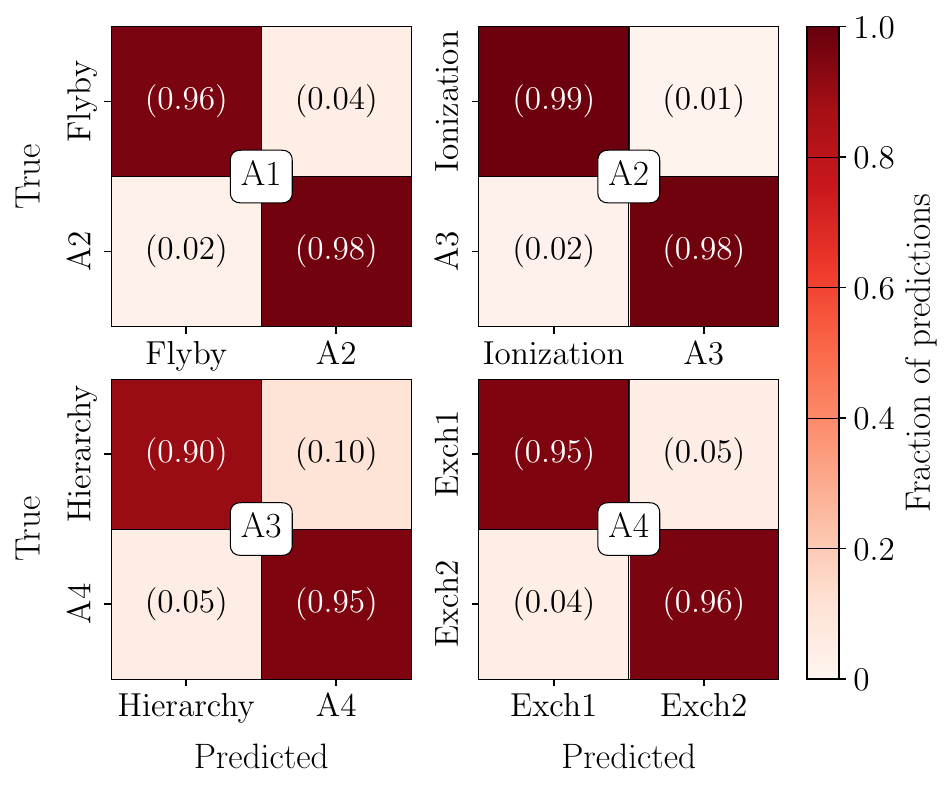}
	\caption{Confusion matrices for the cascaded binary classifiers (Models A1 to A4). The color intensity reflects the fraction of predictions per true class, with each cell displaying the corresponding fraction. Diagonal cells represent correct predictions, while off-diagonal cells indicate misclassified cases. The sequential classification scheme is illustrated in Fig.~\ref{fig:model_diagram}. Exch1 and Exch2 denote exchange1 and exchange2, respectively.}
	\label{fig:confusion_matrices_A}
\end{figure}

\begin{figure}[h!]
	\centering
	\includegraphics[width=\linewidth]{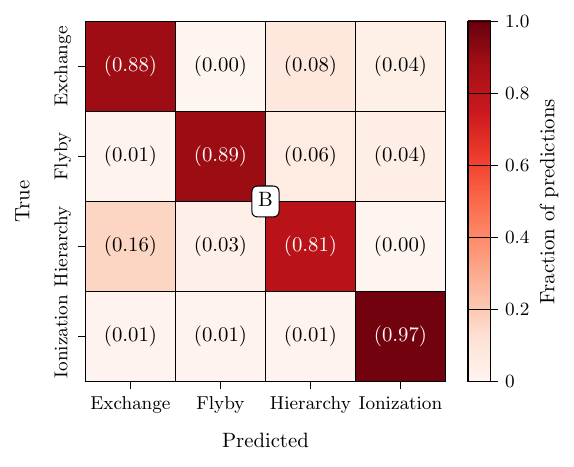}
	\caption{Confusion matrix for Model B.}
	\label{fig:confusion_matrix_B}
\end{figure}

Figures~\ref{fig:confusion_matrices_A} and~\ref{fig:confusion_matrix_B} present the confusion matrices for the cascaded models and Model B, respectively. The former achieve near-perfect accuracy, with Model A2 reaching $99\%$ correct predictions and Model A3 performing worst, correctly identifying $90\%$ of hierarchies. In contrast, Model B underperforms, with per-class accuracies of $89\%$ for flybys, $97\%$ for ionizations, $88\%$ for exchanges, and $81\%$ for hierarchies.

To identify the sources of prediction failures, we compared the mean standardized feature values, $\Delta\mu$, between misclassified and correctly classified cases for each model. For Model A1, misclassified events are characterized by elevated $\Theta$ ($\Delta\mu = 0.450$), and reduced $\mathcal{R}$ ($\Delta\mu = -0.411$). Misclassifications in Model A2 are driven primarily by lower $\mu_{\rm a}$ ($\Delta\mu = -0.501$), followed by higher $\tau$ ($\Delta\mu = 0.348$) and $m_2$ ($\Delta\mu = 0.254$). Similarly, Model A3 shows the largest disparity for $\mu_{\rm a}$ ($\Delta\mu = -0.426$), with additional contributions from $\delta_{\rm m}$ ($\Delta\mu = 0.352$) and $\tau$ ($\Delta\mu = 0.347$). For Model A4, confusing the exchange subtypes are most strongly driven by higher $\mathcal{L}$ ($\Delta\mu = 0.504$), with relatively smaller contributions from $q_2$ ($\Delta\mu = 0.256$) and $v$ ($\Delta\mu = 0.189$). Model B exhibits the largest overall feature disparities, with $\Theta$ ($\Delta\mu = 0.541$) and $\tau$ ($\Delta\mu = 0.446$) being the primary drivers of misclassification, followed by a smaller contribution from $\mu_{\rm a}$ ($\Delta\mu = -0.371$).

\subsection{Speed benchmark and generalization}

Our models make predictions significantly faster than direct $N$-body simulations performed with the REBOUND IAS15 integrator. Models A2 and A4 predict $50\,000$ new binary-single encounters in approximately $0.35$ seconds, outperforming full $N$-body integrations (which require roughly $100$ seconds for the same task) by a factor of nearly $300$. The slowest of our classifiers, Model B, completes the same task in $6$ seconds, still more than $15$ times faster than the direct $N$-body method.

Unlike in our previous study \citep{Farhani2026} on the ML prediction of binary formation from three-body encounters, however, the models presented here failed to generalize to new data, with accuracy dropping below $10\%$. The potential causes and implications of this limitation are discussed in the following section. 

\section{Discussion}
\label{sec:discussion}

Our results show that ML models can learn the nonlinear mapping between initial conditions and final outcomes of binary-single encounters with high accuracy. The cascaded framework outperforms a single multi-class model by decomposing the task into physically motivated binary decisions, allowing each classifier to specialize in a simpler dynamical regime. This approach is both a known ML strategy for reducing multi-class complexity \citep{Dietterich1995, Rifkin2004} and physically motivated, as the final outcome results from successive transitions governed by distinct combinations of energy and angular momentum \citep{Hut1983a, Heggie2003, Valtonen2006}. The outstanding performance of Model A2, along with the high accuracy of Model B for ionization events despite their rarity ($0.8\%$ of the dataset), suggests that ionization occupies a well-defined and distinguishable region of feature space. This is further reflected in the relatively small size of the model file ($6$ MB). As a reason, binary disruption is governed primarily by the encounter energy budget. Ionization occurs only when sufficient orbital binding energy is transferred to unbind the binary, making energetic considerations more decisive than encounter geometry. The dominance of binary type $\mathcal{B}$ as the top feature in Model A2 further supports this interpretation.

Hierarchies are the most challenging outcomes to predict across both the cascaded and multi-class models. This complexity may stem, in part, from the broad definition adopted in this study, which encompasses a wide range of configurations, including cases where the third body is captured by the only bound pair, as well as systems with two or three bound pairs. Moreover, stability assessments \citep{Mardling2001, Vynatheya2022} indicate that all such cases are transient, reinforcing the ambiguity of this class. A dedicated investigation into hierarchies and stability lies beyond the scope of the present work. Nevertheless, both Model A3 and Model B achieve accuracies well above the random-guessing baseline of $0.25$ for these outcomes, suggesting that a substantial portion of the underlying dynamics is captured by the models.

The feature importance analysis reveals distinct patterns across classifiers. Models A1 to A3 are each dominated by a single feature, indicating that their respective classification tasks are governed by one primary physical process. For A1, the encounter timescale ($\tau$) is most influential, consistent with flybys being rapid passages with minimal energy exchange \citep{Hut1983a, Heggie2003}. Models A3 and A4 share the same leading features, $q_3$ and $\Theta$, suggesting that hierarchy and exchange formation are governed by closely related dynamical processes. In contrast, the multi-class model distributes importance more evenly, reflecting the need to simultaneously resolve multiple competing processes across the full parameter space.

The features driving misclassifications differ from those most important ones for predictions, suggesting that errors arise from secondary parameters that become relevant near decision boundaries, consistent with the chaotic nature of resonant scattering \citep{Valtonen2006}. The overall high accuracy indicates these ambiguous regions occupy only a small fraction of the parameter space.

The substantial speed-up achieved by the ML models demonstrates their potential as surrogate predictors for binary-single scattering outcomes, offering an attractive alternative to direct $N$-body integrations in applications where only the encounter final state is required. This can include focused studies on binary survival (flybys), complete disruption (ionization), and exchanges. Rather than replacing high-precision integrators, such models can complement them by rapidly exploring large parameter spaces or serving as components of population synthesis and cluster evolution calculations. A similar framework could be developed for existing dynamical codes, including \textsc{NBODY6++GPU} \citep{Wang2015} and \textsc{PeTar} \citep{Wang2020}, by training surrogate models on code-specific encounter datasets. Although our models do not generalize beyond their training distribution, this reflects the well-known dependence of supervised learning on the training data \citep{Segovia2025} rather than a methodological flaw. The present work thus establishes a general strategy for constructing accurate, domain-specific surrogate models adaptable to a wide range of astrophysical few-body applications.

The failure to generalize beyond the training distribution can be attributed to several fundamental factors inherent to the nature of the three-body problem and supervised learning. First, the chaotic dynamics of resonant binary-single encounters produce fractal basin boundaries in the initial condition space, meaning that infinitesimally small perturbations can lead to qualitatively different outcomes \citep[see, e.g.,][]{Assis2014, Zotos2016}. This intrinsic sensitivity makes the mapping between initial conditions and final states inherently discontinuous in certain regions, posing a fundamental challenge for any deterministic predictor. Second, the training data, despite its large size, necessarily represents a finite sampling of a high-dimensional parameter space. Regions that are sparsely sampled or lie near dynamical transition boundaries are unlikely to be adequately represented. Third, the downsampling procedure, while essential for class balance, alters the effective distribution on which the model is trained, potentially creating decision boundaries that are optimized for the balanced distribution rather than the realistic one. Finally, the complexity of the underlying physics, involving coupled energy exchange, angular momentum redistribution, and long-term stability, likely requires a much larger and more diverse training set to achieve genuine generalization across the full parameter space.

\section{Future work and limitations}
\label{sec:future_works}

Several limitations of the present study should be acknowledged. First, our objective was to test a general ML framework for binary-single scattering, rather than to develop a definitive solution applicable to all possible configurations. Consequently, ambiguous outcomes such as hierarchies remain insufficiently characterized, as they encompass a broad range of transient and metastable states that were not treated with the level of detail they require. Second, we assumed isolated encounters for all simulations, neglecting the external gravitational potential present in realistic environments, such as the tidal field of a globular cluster or the potential of a supermassive black hole \citep{Trani2024}. Third, the initial parameter ranges and preprocessing criteria adopted in this work were intentionally broad, which may not be optimal for specialized applications. Studies focused on specific environments, such as black hole binaries in globular or nuclear star clusters, may yield improved performance by tailoring the training distribution to the relevant parameter space. Finally, while the chaotic boundary analysis proved valuable for identifying challenging regions, our exploration of this aspect remains shallow and primarily serves to demonstrate the utility of the approach rather than to exhaustively map the fractal structure of the parameter space.

Future work should address these limitations by refining the classification of hierarchies, incorporating more detailed stability criteria, accounting for external potentials, and targeting parameter regimes relevant to specific astrophysical contexts. Moreover, existing $N$-body and Monte Carlo codes, such as NBODY6++GPU, PeTar, or MOCCA/CMC \citep{Hypki2013, Giersz2013, Fregeau2004}, could adopt the same strategy to build surrogate models tailored to their respective assumptions and applications, thereby improving computational efficiency while maintaining accuracy in the regimes they are designed to handle.

\section{Summary and conclusion}
\label{sec:conclusions}

We generated $5.8 \times 10^6$ binary-single scattering simulations using REBOUND's IAS15 integrator. Following feature construction and hyperparameter optimization, we adopted two ML strategies: a cascaded sequence of binary classifiers (Models A1 to A4) and a single multi-class classifier (Model B). The cascaded approach outperforms Model B across all metrics and classes, even though both achieve evaluation metrics exceeding $0.90$. Low Brier scores and ECE values indicate that the models are well calibrated. Ionization is the easiest outcome to predict in both strategies, while hierarchy proves the most challenging.

Feature importance analysis revealed physically interpretable patterns aligned with theoretical expectations. The encounter timescale $\tau$ dominates the flyby versus disruption decision (Model A1), the binary type $\mathcal{B}$ drives ionization predictions (Model A2), and the mass ratio $q_3$ together with the Sofronov number $\Theta$ governs the exchange and hierarchy distinctions (Models A3 and A4). Misclassification analysis further shows that the regions in parameter space responsible for confusion are distinct from those where predictions are reliable.

Our models are up to $300$ times faster than direct $N$-body integrations. However, they failed to generalize to new datasets, with accuracy dropping substantially, reflecting the dependence of supervised learning on the training distribution and the intrinsic difficulty of chaotic dynamics.

In conclusion, the cascaded binary strategy provides a flexible and interpretable framework for surrogate modeling of binary--single scattering. Despite generalization limitations, the considerable speed-up makes these models valuable for rapid parameter-space exploration and population synthesis. This work establishes a proof of concept that ML can capture dominant chaotic dynamics, while highlighting the need for careful data curation and uncertainty quantification.

\begin{acknowledgements}	
	We acknowledge the use of ChatGPT (OpenAI) and DeepSeek (DeepSeek AI) for language refinement, code debugging, and literature search during the preparation of this manuscript. The core scientific ideas, methodology, numerical simulations, data analysis, and interpretation of results were carried out by the authors without AI assistance. All AI-assisted outputs were reviewed, verified, and the authors take full responsibility for the content.	
\end{acknowledgements}

\bibliographystyle{aa}
\bibliography{references}

\end{document}